\documentclass[11pt,a4paper]{article}
\pdfoutput=1
\usepackage{jstyle}

\usepackage{cite}

\newcommand{\la}{\langle}
\newcommand{\ra}{\rangle}

\newcommand{\be}{\begin{equation}}
\newcommand{\ee}{\end{equation}}
\newcommand{\bs}{\begin{subequations}}
\newcommand{\es}{\end{subequations}}
\newcommand{\ba}{\begin{aligned}}
\newcommand{\ea}{\end{aligned}}
\newcommand{\bg}{\begin{gathered}}
\newcommand{\eg}{\end{gathered}}
\newcommand{\bfi}{\begin{figure}}
\newcommand{\efi}{\end{figure}}

\def\a{\alpha}
\def\b{\beta}
\def\g{\gamma}

\def\D{\Delta}

\def\h{\eta}
\def\th{\theta}
\def\Th{\Theta}

\def\k{\kappa}
\def\l{\lambda}
\def\L{\Lambda}
\def\m{\mu}
\def\n{\nu}

\def\p{\pi}

\def\s{\sigma}
\def\S{\Sigma}

\def\f{\phi}

\def\vf{\varphi}
\def\c{\chi}

\def\o{\omega}
\def\O{\Omega}

\def\cJ{{\cal J}}

\def\cO{{\cal O}}
\def\cP{{\cal P}}

\def\cT{{\cal T}}

\def\fso{{\mathfrak {so}}}
\def\fiso{{\mathfrak {iso}}}

\def\fheis{{\mathfrak {heis}}}

\def\fg{{\mathfrak g}}

\def\fu{{\mathfrak u}}

\def\bbR{{\mathbb R}}

\def\bbY{{\mathbb Y}}

\def\dd{{\rm d}}
\def\Ad{{\rm Ad}}
\def\ad{{\rm ad}}

\renewcommand{\span}[1]{{\rm Span}\left\{#1 \right\}}

\title{\centering
Ab Initio Construction of  \\ Poincar\'e and AdS Particles}

\author[a,b]{TaeHwan Oh}

\affiliation[a]{Department of Physics and Research Institute of Basic
  Science, \\ Kyung Hee University, Seoul 02447, Korea}
\affiliation[b]{Quantum Universe Center, Korea Institute for Advanced Study, Seoul 02455, Korea}

\emailAdd{hepthoh@kias.re.kr}

\abstract{
We study the construction of a manifestly covariant worldline action 
from a coadjoint orbit. 
A coadjoint orbit is a submanifold in the dual vector space of a Lie algebra,
generated by coadjoint actions.
Since a coadjoint orbit is a symplectic space, 
we derive the worldline particle action from the symplectic two-form.
One subtlety in formulating worldline particle actions from coadjoint orbits
is the choice of a coordinate system 
that clearly illustrates physical properties of the particles.
We introduce Hamiltonian constraints derived from the defining conditions of the isometry.
This allows us to write a manifestly covariant worldline action.
We demonstrate our method for both massive and massless particles 
in Minkowski and AdS spacetime.
}

\begin{document}

\maketitle

\section{Introduction}
\label{sec:intro}

There have been many attempts to formulate spinning particles since the 1920s,
when a classical description of an electron was first proposed \cite{Frenkel:1926zz}.
Modern approaches based on supersymmetry were developed in the late 1970s
\cite{Gershun:1979fb,Casalbuoni:1976tz,Berezin:1976eg,Brink:1976sz,Barducci:1976qu}
(see also \cite[Section 1]{part1} for additional references).
Over the past century, worldline particle actions have played important roles
in various areas of physics, e.g., 
\cite{Bastianelli:2008vh,Bastianelli:2023oyz, Comberiati:2022cpm,Rempel:2015foa, Schubert:2023bed}.
Despite their significance, 
a systematic method for constructing worldline actions
for various particle species in different spacetimes
remains underdeveloped.

\medskip

In our previous paper \cite{part1}, we introduced a method for constructing actions for relativistic spinning particles
based on the orbit method, which we will employ in this work.
The orbit method has been studied since the 1960s and 1970s 
\cite{kirillov1962unitary,kirillov2004lectures,auslander1967quantization,Kostant:1969zz,auslander1971polarization,souriau1970structure}
as a means to obtain unitary irreducible representations
through the quantization of coadjoint orbits, which are symplectic.
These efforts provided inspiration for the development of worldline actions for particles 
\cite{Duval:2014ppa, Rempel:2015foa,Andrzejewski:2020qxt, part1}. 

A significant challenge in constructing worldline actions
using the orbit method 
is the choice of an appropriate coordinate system of the orbit.
We take the covariance of the coordinate system as the criterion for appropriateness.
To formulate the worldline action in a manifestly covariant way,
Hamiltonian constraints must be incorporated.
For instance, as we will see later, a Poincar\'e scalar particle action obtained from
a scalar coadjoint orbit is written as 
\be 
	S = \int \la m\cP^0 \,,\, g^{-1}\dd g  \ra\,,
\ee
where $g$ is an $ISO(1,d-1)$ element and 
$\cP^0$ is the dual of Lie algebra generator $P_0$.
A manifestly covariant worldline action for a scalar particle
is given as
\be
	S[x\,, p\,,  e] = \int\, p_\m \dd x^\m+\frac e2 \left( p^2+m^2 \right)\,.
\ee
This action contains a Hamiltonian constraint $p^2+m^2\approx 0$.
We will see that the constraint can be obtained from the
defining condition of the $SO(1,d-1)$ group.

\medskip

In our previous paper \cite{part1}, we explained the general framework for 
deriving worldline actions from the coadjoint orbits.
The main objective of this paper is to demonstrate this method explicitly, 
as introduced in the previous paper,
for both massive and massless particles in Minkowski and AdS spacetime.
In Section \ref{sec:review}, we explain how to construct worldline particle actions
of $ISO(1,d-1)$ and $SO(2,d-1)$ isometries using the orbit method.
To this end, we begin the section with a pedagogical review of the orbit method.
In Section \ref{sec:coadj orbit}, 
we  apply the method to analyse these particles.
Finally, in Section \ref{sec:action}, 
we show how the worldline actions for these particles
are derived by applying the techniques developed in the earlier sections.
In Appendix \ref{sec:proof}, we provide a proof 
of a mathematical property used in Section \ref{subsec:action}.

\section{Covariant Action from the Coadjoint Orbit}
\label{sec:review}

In this section, we will briefly review how to construct a covariant worldline action from a coadjoint orbit.
For more detailed discussions of the orbit method 
and construction of worldline actions from coadjoint orbits,
see  e.g. \cite{souriau1970structure,kirillov2004lectures,kirillov2012elements, vogan1998method, Oblak:2016eij,part1}. 
	
	\subsection{Review of the Coadjoint Orbit}
	For a Lie group $G$, there is a left-invariant vector space called Lie algebra $\fg$, 
	and  a Lie algebra dual space $\fg^*$,
	which forms a natural pairing $\la \cdot \,, \cdot \ra$ with $\fg$ such that
	\be  \label{eq:dual space}
		\la \cT^a\,, T_b \ra = \delta^a_b\,,
	\ee
	where $T_a$ and $\cT^a$ are generators of $\fg$ and $\fg^*$, respectively.
	Note that when $G$ is a simple group, we can identify $\fg^* \simeq \fg$
	since there exists a non-degenerate bilinear form.
	
	A group action on $\fg^*$, denoted by $\Ad^*_g$,
	is called a coadjoint action, 	
	and is defined as the dual of the adjoint action:
	\be \label{eq:coadjoint action}
		\la \Ad_g^*\f\,, X \ra = \la \f \,, \Ad_{g^{-1}} X \ra \,,
	\ee
	where $\f\in \fg^*\,,\ X\in\fg\,,$ and $g\in G$.
	A coadjoint orbit $\cO^\f$ is the submanifold generated by coadjoint actions on $\f$, such that
	\be \label{eq:coadjoint orbit}
		\cO^\f=\{\vf\in \fg^*\, |\,   \Ad^*_g\f=\vf\,,\ g\in G \}\,,
	\ee
	and the stabiliser $G^\f$ is the isotropic subgroup for the coadjoint action:
	\be \label{eq:stabiliser}
		G^\f=\{g\in G\, |\,  \Ad^*_g\f=\f \}\,.
	\ee
	Since different $\vf$'s within $\cO^\f$ belong to the same equivalence class 
	with equivalence relation $\f \sim \vf=\Ad_g^*\f$,
	a specific vector $\f$ can serve as a representative vector of $\cO^\f$. 
	
	One immediately finds that the coadjoint action $\ad^*_X$ by a Lie algebra vector $X\in\fg$ 
	is given by
	\be \label{eq:inf coadj action}
		\la \ad_X^*\f\,, Y \ra = -\la \f \,, \ad_{X} Y \ra = -\la \f \,, [X\,, Y]\ra \,.
	\ee
	The tangent space of the coadjoint orbit is given by
	\be
		T_\vf\cO^\f=\{X\in \fg\ |\,  \ad^*_X \f =\vf\,,\ \vf\in \cO^\f \}\,,
	\ee
	and the stabiliser subalgebra is
	\be\label{eq:inf stab}
		\fg^\f  =\{X\in G\, | \, \ad^*_X \f =0 \}\,.
	\ee
	From \eqref{eq:inf coadj action}, any element $X\in \fg^\f$ must satisfy
	\be\label{eq:lie alg stab}
	\la \f\,, [X,Y]\ra=0\,.
	\ee
	for all $Y\in \fg$.
	
	Note that $G$ can be regarded as a principal $G^\f$--bundle over $\cO^\f$, with projection $\p_\f$:
	\be \label{eq:projection} \ba
		\p_\f\,:\ 	G &\rightarrow \cO^\f\,, \\
				g &\mapsto \p_\f(g)=\Ad^*_g\f\,,
	\ea \ee 
	then $\cO^\f$ is a homogeneous space, given by
	\be \label{orbit hom}
		\cO^\f\simeq G/ G^\f\,.
	\ee
	
	One important feature of coadjoint orbits is 
	their nature as a symplectic space, 
	that is, $\cO^\f$ is always an even-dimensional manifold
	equipped with a non-degenerate, closed two-form $\o$,
	known as the KKS(Kirillov-Kostant-Souriau) two-form.
	KKS two-form at $\vf \in \cO^\f$ is denoted by $\o_\vf$
	and is given by
	\be\label{eq:kks form}
		\o_\vf\left(\ad^*_X\vf \,, \ad^*_Y\vf\right)=\la \vf\,, [X,Y]\ra\,.
	\ee
	Using the projection operation in \eqref{eq:projection},
	the pull-back of the KKS two-form $\o$ onto group $G$ is given by
	\be \label{eq:KKS pull-back}
		\O\equiv \p^*_\f(\o)= \la \f\,, [\Th(X),\Th(Y)] \ra  =-\dd\la \f \,, \Th \ra\,,
	\ee
	where $\Th$ is a $\fg$-valued left-invariant one-form on $G$,
	which is expressed as $\Th=g^{-1}\dd g$.
	We used the Maurer-Cartan equation in the last equality in \eqref{eq:KKS pull-back}. 

	Since the two-form $\O$ is an exact form on $G$, we can find a symplectic potential $\th$ as
	\be
		\th=\la \f \,, \Th \ra\,,
	\ee
	so a worldline action is written in terms of the symplectic potential by
	\be \label{eq:worldline}
		S=\int_\g \theta=\int_\g \la \f\,, \Th \ra = \int_\g \la \f\,, g^{-1}\dd g \ra\,,
	\ee
	where $\g\in G$ is a path lying in the group. 
	If one takes $G$ as an isometry group, 
	then the coadjoint orbit $\cO^\f$ can be interpreted 
	as the phase space of a particle,
	with its physical information encoded in $\f$.
	Consequently, the action \eqref{eq:worldline}, derived from the coadjoint orbit,
	can be understood as the worldline action for a relativistic particle.
	
	\medskip
	
	\subsection{Manifestly Covariant Worldline Action} \label{subsec:action}
	So far, we discussed how to derive a worldline action from a coadjoint orbit.
	However, a subtlety arises when choosing an appropriate coordinate system for the orbit.
	In our approach, we adopt covariance as the criterion for selecting a suitable coordinate system.
	Manifestly covariant worldline actions are formulated
	by incorporating Hamiltonian constraints into the action.
	As we will show later, these constraints are derived from the defining conditions
	of the isometry.
	
	Manifestly covariant worldline actions for 
	various groups, including classical Lie groups and semi-direct groups,
	were formulated systematically in our previous paper  \cite[Section 4]{part1}.
	In the remainder of this section, 
	we will focus on the cases of $SO(2,d-1)$ and $ISO(1,d-1)$,
	which are isometry groups of the $d$-dimensional AdS and flat spacetime, respectively.
	
		\subsubsection{Covariant Formulation of $SO(2,d-1)$ action} \label{subsec:cov form so}
		For a matrix group, \eqref{eq:worldline} can be expressed in terms of matrices.
		Thus, the worldline action for $SO(2,d-1)$ can be rephrased as
		\be \label{eq:worldline 1}
			S[X]=\int \tr(\f\, X^{-1}\dd X)=\int \f^a{}_b\, (X^{-1})^b{}_c\, \dd X^c{}_a\,,
		\ee
		for $X\in SO(2,d-1)$. 
		To make a manifestly covariant worldline action, 
		we introduce the defining condition of $SO(2,d-1)$, that is, $X^T\h X=\h$,
		where $\h$ is the metric $\h=(-\,,-\,,+\,,+\,,\ldots)$,
		then the action \eqref{eq:worldline 1} becomes
		\be
			S[X,A] 	=\int  \left[ \f^a{}_b\, (X^{-1})^b{}_c\, \dd X^c{}_a+ 
					A^{ab} \big( X^c{}_b\, \h_{cd}\,  X^d{}_{a} - \h_{ba}  \big ) \right]\,,
		\ee
		where $a\,,b\,,\cdots=0'\,,0\,,1\,,\cdots\,,d-1$ are $(d+1)$-dimensional indices, 
		and $A^{ab}$ is a Lagrange multiplier.
		Using the constraint, $X^{-1}$ can be substituted by $\h\, X^T\h$, so
		\be \ba
			S[X,A]	&=\int  \left[ \f^a{}_b\, \h^{bc}\, X_{dc} \h^{de}\, \dd X_{ea}+ 
					A^{ab} \big( X^c{}_b\, \h_{cd}\,  X^d{}_{a} - \h_{ba}  \big ) \right] \\
					&=\int  \left[ \f^{ab}\, X^c{}_b\, \dd X_{ca}+ 
					A^{ab} \big( X^c{}_b\, \h_{cd}\,  X^d{}_{a} - \h_{ba}  \big ) \right]\,.
		\ea \ee
		Since $\f^{ab}$ is an $\fso(2,d-1)^*$ element, it is anti-symmetric.
		Moreover, $\f^{ab}$ can be expressed as a symplectic matrix through a 
		congruent transformation.
		Let $\rank\f=M\leq d+1$,
		\footnote{Since $\f$ is an anti-symmetric and non-degenerate matrix, $M$ must be even.}
		 then $\f$ can be written as
		\be\label{eq:poin congruent}
			\f^{ab}=T^a{}_\a \O^{\a\b} T^b{}_\b\,,
		\ee
		where $\a\,,\b=1\,,\cdots\,,M$ are symplectic indices,
		$\O$ is an $M\times M$ symplectic matrix,
		and $T$ is an $M\times (d+1)$ matrix.
		Note that we can always find a matrix $T$ 
		which transforms the symplectic matrix to
		any anti-symmetric matrix 
		(see Appendix \ref{sec:proof} to find a proof of this statement.).

		Once we define the variables as
		\be \label{eq:redef var}
			Y^c{}_\a = X^c{}_{a} T^a{}_\a\,,\quad A^{\a\b}=A^{ab}\, T^b{}_\b\, T^a{}_\a\,,
			\quad \vf_{\b\a}=T^a{}_\a\, T_{a\b}\,,
		\ee
		then the worldline action can be rewritten as
		\be\label{eq:hom action}
			S[Y,A] = \int \left[ \O^{\a\b} Y^c{}_\b\,\dd Y_{c\a} +
			A^{\a\b}\big(  Y^c{}_\b Y_{c\a} - \vf_{\b\a}\big) \right]\,,
		\ee
		after integrating out the spurious variables $X^a{}_i$ such that $X^a{}_i\,T^i{}_\a=0$
		for all $\a$ since they are non-dynamical.
		
		It is worth emphasising that 
		the form of the action \eqref{eq:hom action} is universal,
		and it is sufficient to identify $\vf$ from the representative $\f$ of the orbit
		to obtain a manifestly covariant worldline particle action
		associated with the orbit $\cO^\f$
		
		\subsubsection{Covariant Formulation of $ISO(1,d-1)$ action}
		For Poincar\'e group $ISO(1,d-1)$,
		a representative vector $\f$ consists of
		an inhomogeneous part $\f_I\in \bbR^{1,d-1}{}^*$ and 
		a homogeneous part $\f_H\in \fso(1,d-1)^*$
		(i.e. $\f=\f_I+\f_H$).
		Let $\L$ be an $SO(1,d-1)$ element and $x$ be a $\bbR^{1,d-1}$ element, 
		then the worldline action is written as follows:
		\be \label{eq:inhom action 1}
			S[x\,, \L]=\int \tr \big[ \f_I \L^{-1} \dd x + \f_H \L^{-1} \dd\L   \big]\,.
		\ee
		As in Section \ref{subsec:cov form so}, we introduce a Hamiltonian constraint 
		as the defining conditions of $SO(1,d-1)$, 
		then $\L$ can be treated as a $GL(d,\bbR)$ element.
		Using the Hamiltonian constraint, $\L^{-1}$ can be replaced by $\h\, \L^T\h$, 
		and we find
		\be \label{eq:inhom action int} \ba
			S[x\,, \L\,, A] &= \int \left[(\f_I)_a\, (\L^{-1})^a{}_b\, \dd x^b+ (\f_H)^a{}_b (\L^{-1})^b{}_c\, \dd \L^c{}_a + A^{ab}\big( \L_{cb}\,\L^{c}{}_a-\h_{ba} \big)   \right]\,, \\
						&=\int \left[(\f_I)^a\, \L_{ba}\, \dd x^b+ (\f_H)^{ab}\, \L^c{}_b \dd \L_{ca} + A^{ab}\big( \L_{cb}\L^{c}{}_a-\h_{ba} \big)   \right] \,,
		\ea \ee
		where $a,b=0,1\,,\cdots\,,d-1$.
		As discussed in Section \ref{subsec:cov form so} and Appendix \ref{sec:proof},
		$\f_H$ can be expressed using a symplectic matrix by congruence transformation
		$\f_H{}^{ab}=T^a{}_\a\, \O^{\a\b}\, T^b{}_\b$,
		where $\a,\b=1\,,\cdots\,, M$ are $Sp(M,\bbR)$ indices 
		for $M=\rank\f_H\leq d$.
		
		Before redefining the variables along the transformation, 
		the Lagrange multiplier $A^{ab}$ can be decomposed into 
		$A^{ab}= \f_I{}^a \f_I{}^b  A + \f_I {}^{(a}  A^{b)}+A^{ab}$,
		then the action \eqref{eq:inhom action int} becomes
		\be \ba \label{eq:inhom action}
			S[x\,, \L\,, A] = \int \Big[p_a\, \dd x^a &+ \O^{\a\b}\, \S_{c\b}\, \dd \S^c{}_\a + A\big(p^a p_a - \vf_M \big) \\
			   &+A^\a\big(p_a \S^a{}_\a - \vf_I{}_\a \big) +A^{\a\b}(\S_{c\b}\S^c{}_\a-\vf_H\, {}_{\b\a}) \Big]\,,
		\ea \ee
		after integrating out spurious variables like $\L^a{}_i$ 
		such as $\L^a{}_i\,T^i{}_\a=0$ for all $\a$, 
		and $(\f_I)^i=0$, since they are non-dynamical.
		We redefine the variables in \eqref{eq:inhom action} as 
		\be \bg \label{eq:subst poin}
		p_a=(\f_I)^b \L_{ab}\,,\quad \S^c{}_\a=\L^c{}_\a\,, \\
		\vf_M=\f_I{}_a \f_I{}^a \,, \quad \vf_I{}_\a=\f_I{}_a T^a{}_\a \,, \quad \vf_H{}_{\a\b}=T^a{}_\a\, T^b{}_\b\, \h_{ab}\,.
		\eg \ee
		Similar to the $SO(2,d-1)$ case,
		the action \eqref{eq:inhom action} is universal.
		Once one extracts $\vf_M\,, \vf_I\,, $ and $\vf_H$ 
		from the original representative vector $\f$ of the coadjoint orbit $\cO^\f$, 
		one can obtain a manifestly covariant worldline action of the particle
		corresponding to $\cO^\f$.

\section{Stabiliser Analysis of Coadjoint Orbits}
\label{sec:coadj orbit}

In this section, we study the coadjoint orbits of 
Poincar\'e and AdS by analysing their stabiliser algebras,
which are determined by solving systems of linear equations. 

For the sake of simplicity, 
we will focus on the three specific cases:
Poincar\'e massive particles, represented by $\f=m\,\cP^0+s\,\cJ^{12}$,
Poincar\'e massless particles, represented by $\f=E\,\cP^++s\,\cJ^{12}$,
and AdS particles represented by $\f=m\,\cJ^{0'0}+s\,\cJ^{12}$.

	\subsection{Stabiliser Algebra of Poincar\'e Particles}
	
	Since representative vectors of $\fiso(1,d-1)$ consist of
	inhomogeneous parts $\f_I$ and homogeneous parts $\f_H$,
	they can be classified according to these components.
	The inhomogeneous part $\f_I$ characterises the mass type of particle.
	For instance, $\f_I=m\,\cP^0$ corresponds to a massive particle,
	$\f_I=E\,\cP^+$ to a massless particle,
	and $\f_I=\m\,\cP^1$ to a tachyonic particle.
	The homogeneous part, $\f_H$, on the other hand,  
	determines the spin of the particles from the Young diagram,
	except in the case of a continuous spinning particle, $\f=E\,\cP^++\s\,\cJ^{-1}$.
	For instance, when
	\be
		\f_H=s_1\,\cJ^{12}+s_2\,\cJ^{34}+\cdots\,,
	\ee
	the spin structure is given by the Young diagram
	\footnote{The components of $\f_H$ can take only integer values 
	due to the consistency of the quantization condition.},
	\be
		\bbY= (s_1\,,s_2\,,\ldots)\,.
	\ee
	In what follows, we restrict our focus to massive and massless particles.
	Remark that this classification of $\f$ is in accordance 
	with the Wigner's classification \cite{Wigner:1939cj, Bargmann:1948ck}.
	
	To identify the stabiliser algebra, 
	we use the general expression from \eqref{eq:lie alg stab}
	and parametrise:
	\be
		\f=\a_{a}\cP^a+\b_{ab}\cJ^{ab}\,,\quad X=A^a\, P_a+B^{ab}\, J_{ab} \,,\quad Y=C^a\, P_a + D^{ab}\, J_{ab}\,.
	\ee
	By imposing  
	\eqref{eq:inf stab}\footnote{We used commutator relations 
	between $\fiso(1,d-1)$ generators are given by
	\be \bg
	{}[J_{ab}\,,J_{cd}]=\h_{ac} J_{bd}-\h_{ad} J_{bc}-\h_{bc} J_{ad}+\h_{bd} J_{ac}\,, \\
	[P_a\,, J_{cd}]=\h_{ac}P_{d}-\h_{ad}P_c\,,\quad [P_a\,, P_b]=0\,.
	\eg \ee}
	, we find
	\be\ba \label{eq:paring stab}
		\la \f \,, &[X\,,Y] \ra \\
		&= C^c (2B^{ab}\, \h_{ac}\, \a_b) 
				+ D^{cd}\big[2B^{ab}\, (\h_{ac}\, \b_{bd}-\h_{ad}\, \b_{bc})-A^a\, (\h_{ac}\, \a_d-\h_{ad}\, \a_c)\big ]\,.
	\ea\ee
	This must vanish for arbitrary $C^c$ and $D^{cd}$,
	finding the linear equations
	\bs \label{eq:stab eqn}
		\be \label{eq:stab eqn1}
			B^{ab}\, \h_{ac}\, \a_b=0\,,
		\ee
		\be \label{eq:stab eqn2}
			2B^{ab}\, (\h_{ac}\, \b_{bd}-\h_{ad}\, \b_{bc})-A^a\, (\h_{ac}\, \a_d-\h_{ad}\, \a_c)=0\,.
		\ee
	\es	
	In conclusion, once the data $\{\a_a\,,\b_{ab}\}$ are determined from the representative,
	the stabiliser subalgebra can be determined from the equation \eqref{eq:stab eqn}.

	Let us now consider the case of massive scalar particles.
	Such a particle is described by the following representative vector,
	\be \label{eq:poin scalar rep}
		\f_I=m\,\cP^0\,,
	\ee
	then $\a_0=m$, $\a_i=0$, and $\b_{ab}=0$ for $i=1\,,\ldots\,,d-1$.
	Imposing these data to \eqref{eq:stab eqn1}, we find 
	\begin{itemize}
		\item $B^{a0}\, \h_{ac}\, \a_0=0\, \Rightarrow\, B^{0i}=0\,.$
	\end{itemize}
	Also, we find another result  
	\begin{itemize}
		\item $A^{a}\,\h_{ai}\,\a_0=0\, \Rightarrow\, A^i=0\,,$ 
	\end{itemize}
	from \eqref{eq:stab eqn2}.
	From these two results, 
	we find that the stabiliser algebra of a massive scalar particle $\fg^{\f_I}$ is given by
	\be \label{eq:poin scalar stab alg}
		\fg^{\f_I}=\span{P_0\,,\  J_{ij}}\,,
	\ee
	for $i,j=1\,,\ldots\,, d-1$.
	Since $\fg^{\f_I}$ is isomorphic to $\bbR\oplus \fso(d-1)$,
	it is a $\frac12 (d^2-3d+4)$-dimensional vector space.
	This implies that the coadjoint orbit $\cO^{\f_I}$  of the massive scalar particle 
	has $2(d-1)$-dimension,
	which matches the expected dimension of the classical phase space of a scalar particle.
	
	A massless scalar particle corresponds to the following representative vector:
	\be \label{eq:poin massless scalar rep}
		\f_{I'}=E\,\cP^+\,,
	\ee
	where $X_\pm=X_0\pm X_{d-1}$,
	so that $\a_+=E$, $\a_i=0$, and $\b_{ab}=0$ for $i=-\,,1\,,\ldots\,,d-2$.
	Using these data, we obtain,
	\begin{itemize}
		\item $B^{a+}\,\h_{ac}\,\a_+=0\, \Rightarrow\, B^{+i}=0\,,$
	\end{itemize}
	from \eqref{eq:stab eqn1} and
	\begin{itemize}
		\item $A^a\,\h_{ai}\,\a_+=0\, \Rightarrow \, A^{a}=0\,,$ unless $a=-\,.$
	\end{itemize}
	from \eqref{eq:stab eqn2}.
	Therefore, the stabiliser subalgebra of a massless particle is given by
	\be \label{eq:poin massless scalar stab alg}
		\fg^{\f_{I'}}=\span{P_-\,,J_{-\a}\,, J_{\a\b}}\,,
	\ee
	for $\a\,,\b=1\,,\cdots\,,d-2$.
	The stabiliser subalgebra $\fg^{\f_{I'}}$ is isomorphic to $\bbR\oplus \fiso(d-2)$,
	hence the coadjoint orbit of a massless scalar particle $\cO^{\f_{I'}}$ is also a $2(d-1)$-dimensional phase space.
	
	Note that the two coadjoint orbits $\fg^{\f_I}$ and $\fg^{\f_{I'}}$ have different characteristics.
	The stabiliser of the massive scalar \eqref{eq:poin scalar stab alg} includes its dual generator $P_0$,
	whereas the stabiliser of the massless scalar \eqref{eq:poin massless scalar stab alg} 
	does not include its dual generator $P_+$.
	Mathematically, a coadjoint orbit $\cO^\f$ is called a nilpotent orbit 
	if the condition $\la \f \,, X \ra =0$ holds for all $X\in \fg^\f$.
	If the condition is not satisfied, then it is referred to as a semisimple orbit.
	According to this classification, the coadjoint orbit of the massive scalar particle is semisimple,
	while that of the massless scalar particle is nilpotent.
	
	We now turn to the spinning particle cases. 
	To describe a massive spinning particle,
	we need to extend the representative vector \eqref{eq:poin scalar rep}
	by appending an additional dual vector $\f_H$,
	which should lie in the dual of the little algebra. 
	We choose an additional dual vector $\f_H=\cJ^{12}$
	as a representative for spin,
	in addition to the massive component $\f_I$.
	Therefore, the representative vector for a massive spinning particle is given by
	\be \label{eq:poin spin rep}
		\f=m\,\cP^0+s\,\cJ^{12}\,.
	\ee
	By applying \eqref{eq:stab eqn} with the given data $\a_0=m$, $\b_{12}=s$,
	and setting all other $\a$ and $\b$ to zero,
	we find the solutions from \eqref{eq:stab eqn1},
	\begin{itemize}
		\item $B^{a0}\, \h_{ac}\, \a_0=0\, \Rightarrow\, B^{0a}=0\,,$
	\end{itemize}
	and from \eqref{eq:stab eqn2} for each $(c\,, d)$ pairs
	\begin{itemize}
		\item $(0\,, 1) \quad 2s\, B^{02}-m\, A^1=0\quad
		\Rightarrow A^1=0\,; $
		\item $(0\,, 2) \quad -2s\, B^{01}-m\,A^2=0 \quad
		\Rightarrow A^2=0\,;$
		\item $(0\,, \m)  \quad m\, A^\m=0 \quad
		\Rightarrow A^\m=0\,;$
		\item $(1\,, \m)  \quad s\, B^{2\m}=0 \quad
		\Rightarrow B^{2\m}=0\,;$
		\item $(2\,, \m)  \quad s\, B^{1\m}=0 \quad
		\Rightarrow B^{1\m}=0\,,$
	\end{itemize}
	 where $\m=3\,,\ldots\,, d-1$.
	As a result, the stabiliser subalgebra is given by
	\be \label{eq:poin spin stab alg}
		\fg^{\f}=\span{P_0\,, J_{12}\,, J_{\m\n}}\,,
	\ee
	and $\fg^{\f}$ is isomorphic to $\bbR\oplus \fu(1)\oplus \fso(d-3)$.
	Thus, it is a $\frac12(d^2-7d+16)$-dimensional algebra.
	This means that the coadjoint orbit $\cO^\f$ is a $2(2d-4)$-dimensional phase space.
	
	To describe a massless spinning particle, 
	we introduce an additional dual vector $\f_{H'}$ 
	to the massless representative \eqref{eq:poin massless scalar rep}.
	There are two possibilities for this additional vector, 
	$\f_{H'}=\cJ^{12}$ which corresponds to the massless spinning particle,
	and $\f_{H'}=\cJ^{-1}$,which corresponds to the continuous spinning particle,
	according to Wigner's classification.
	The massless spinning particle is described by the representative vector,
	\be \label{eq:poin massless spin rep}
		\f'=E\,\cP^++s\,\cJ^{12}\,,
	\ee
	hence, $\a_+=E\,,$ $\b_{12}=s\,,$ $\a_-=\a_1=\a_2=\a_i=0$, and $\b_{ij}=\b_{-i}=0$
	for $i\,,j=3\,,\cdots\,,d-2$.
	Using this data, \eqref{eq:stab eqn1} yields:
	\begin{itemize}
		\item $B^{a+}\,\h_{ac}\,\a_+=0\, \Rightarrow\, B^{+-}=B^{+1}=B^{+2}=B^{+i}=0\,,$
	\end{itemize}
	and \eqref{eq:stab eqn2} gives several equations for each $(c\,,d)$ pairs, as follows:
	\begin{itemize}
		\item $(+\,, -) \quad A^+\,\h_{-+}\,\a_+=0\quad
		\Rightarrow A^+=0\,; $
		\item $(+\,,1) \quad 2B^{-2}\,\h_{-+}\,B_{21} + A^1\,\h_{11}\,\a_+=0 \quad
		\Rightarrow s\,B^{-2}+E\,A^1=0\,; $
		\item $(+\,,2) \quad 2B^{-1}\,\h_{-+}\,B_{12} + A^2\,\h_{22}\,\a_+=0 \quad
		\Rightarrow -s\,B^{-1}+E\,A^2=0\,;$
		\item $(+\,, i) \quad A^i\,\h_{ii}\,\a_+=0\quad
		\Rightarrow A^i=0\,; $
		\item $(-\,, 1) \quad 2B^{+2}\,\h_{+-}\,B_{21}=0\quad
		\Rightarrow B^{+2}=0\,; $
		\item $(-\,, 2) \quad 2B^{+1}\,\h_{+-}\,B_{12}=0\quad
		\Rightarrow B^{+1}=0\,; $
		\item $(1\,, i) \quad 2B^{i2}\,\h_{ii}\,B_{21}=0\quad
		\Rightarrow B^{2i}=0\,; $
		\item $(2\,, i) \quad 2B^{i1}\,\h_{ii}\,B_{12}=0\quad
		\Rightarrow B^{1i}=0\,. $
	\end{itemize}	
	Then the stabiliser subalgebra $\fg^{\f'}$ is given by
	\be
		\fg^{\f'}=\span{-E\,J_{-2}+s\,P_1\,,E\,J_{-1}+s\,P_2\,,J_{12}\,,P_-\,,J_{-i}\,,J_{ij}}\,,
	\ee
	which is isomorphic to $(\fheis_2\niplus \fu(1))\oplus \fiso(d-4)$.

	Since the coadjoint orbit is a homogeneous space as defined in \eqref{orbit hom}, 
	its geometric structure is determined by the corresponding stabiliser subgroup.
	At the group level, the stabiliser of a massive scalar particle is 
	\be
		G^{\f_I} = \bbR \times SO(d-1)\,,
	\ee
	so its coadjoint orbit becomes
	\be
		\cO^{\f_I}\simeq  \frac{ISO(1,d-1)}{\bbR \times SO(d-1)} \simeq \bbR^{2(d-1)}\,.
	\ee
	On the other hand, for the massive spinning particle, 
	the coadjoint orbit can be understood
	as a fibre bundle over the base manifold $\cO^{\f_I}$ with a spin fibre,
	\be
		\cO^S \simeq  \frac{SO(d-1)}{SO(2)\times SO(d-3)} 
			\simeq {\rm Gr}_\bbR(d-1,2) \,,
	\ee
	where  $\cO^S$  denotes the phase space for the spin degrees of freedom,
	and ${\rm Gr}_\bbR$ is the real Grassmannian.
	For $d=4$, the spinning particle orbit becomes 
	\be
		\bbR^{6}\times {\rm Gr}_\bbR(3,2)\simeq \bbR^6\times S^2\,,
	\ee
	where the spin phase spaces $S^2$ can be decomposed 
	into various modes with discrete radii. 
	This perspective has been previously discussed in the literature,
	see e.g., \cite{Lyakhovich:1996we,Kuzenko:1994ju}.
	In those works, the worldline actions for 
	the massive spinning particles in $d=4$ flat spacetime
	were constructed geometrically,
	and the quantization of spin was investigated.
	Poincar\'e coadjoint orbits were also classified and analysed in detail in many studies,
	\cite{Rawnsley1975,Carinena:1989uw,Baguis1998,Cushman_2006,hudon2010coadjoint,
	Gracia-Bondia:2017fai,Havlicek:2018tfp,Andrzejewski:2020qxt, Lahlali:2021nrf}.

	We conclude this section with a brief comment 
	on particles with more complicated spin structures. 
	To describe a massive spinning particle, 	
	we append $\f_H=s\,\cJ^{12}$ to the massive scalar representative $\f_I$.
	This procedure can be generalised by adding more vectors to $\f_H$,
	yielding $\f=m\,\cP^0+s_1\,\cJ^{12}+s_2\,\cJ^{34}+\cdots$.
	This configuration describes a particle of a mixed-symmetry field,
	where the spin is given by a Young diagram $\bbY=(s_1\,,s_2\,,\ldots)$.
	The stabiliser subalgebra for such a representative vector is then 
	\be
		\fg^\f=\bbR \oplus \fso(d-1-2n) \oplus \bigoplus_{i=1}^{R} \fu(r_i)\,,
	\ee
	when the Young diagram of this particle is of the form
	\be
		\bbY= (\underbrace{s_1\,,\ldots\,,s_1}_{r_1}\,,\underbrace{s_2\,,\ldots\,,s_2}_{r_2}\,,\ldots)\,.
	\ee

	\subsection{Stabiliser Algebra of AdS Particles}
	\label{subsec:ads orbit}
	
	There are various particle species respecting AdS isometry (see e.g. \cite[Section 6]{part1}), 
	but in this paper we focus on AdS particles with time-like momentum, and space-like spin,
	\be \label{eq:adsparticle}
		\f=m\,\cJ^{0'0}+s\,\cJ^{12}\,.
	\ee
	These particles can be viewed as analogues of Poincar\'e massive particles
	with transvections $\cJ^{0'a}\sim \cP^a$.
	However, this correspondence is subtle for the spinning case, as we will see later.
	For convenience, we refer to \eqref{eq:adsparticle} as AdS spinning particles throughout this paper.
	
	Similarly to the Poincar\'e case, 
	we determine the stabiliser algebra of \eqref{eq:adsparticle}
	by solving the system of linear equations
	\be \label{eq:ads stab eqn}
		B^{ab}\, \h_{ac}\, \b_{bd}-B^{ab}\, \h_{ad}\, \b_{bc}=0\,,
	\ee
	which can be derived by disregarding the inhomogeneous parts of \eqref{eq:stab eqn}.
	
	Let us first examine the scalar case, $s=0$,
	\be
		\f_0=m\,\cJ^{0'0}\,.
	\ee
	Setting $\b_{0'0}=0$ and all other $\b=0$ in \eqref{eq:ads stab eqn},
	we obtain two equations for $C=0\,, 1$ and $D=i$,
	\begin{itemize}
		\item $(0'\,,i)\quad m\, B^{0i}=0 \quad \Rightarrow B^{0i}=0\,;$
		\item $(0\,,i)\quad m\, B^{0'i}=0 \quad \Rightarrow B^{0'i}=0\,,$
	\end{itemize}
	for $i=1\,,\ldots\,,d-1$.
	Hence the stabiliser algebra is
	\be \label{eq:ads scalar stab}
		\fg^{\f_0}=\span{J_{0'0}\,, J_{ij}}\,,
	\ee
	for $i,j=1\,,\ldots\,, d-1$.
	The stabiliser algebra is isomorphic to $\fu(1)\oplus\fso(d-1)$.
	Consequently, the dimension of the corresponding coadjoint orbit $\cO^{\f_0}$
	is $2(d-1)$, which matches the degrees of freedom in the Poincar\'e scalar case.
	Comparing with the Poincar\'e case, 
	the stabiliser of the dimension-independent piece is compact
	since it is generated by $J_{0'0}$,
	and this fact implies that the label $m$ of the orbit $\cO^{\f_0}$ only can take discrete values after quantization.
	The natural interpretation of $m$ is 
	the classical limit of the conformal dimension $\D$,
	and this interpretation justifies the discrete value of $m$ after quantization.
	
	To analyse the stabiliser of  \eqref{eq:adsparticle},
	we consider the case where only $\b_{0'0}=m\,,$ and $\b_{12}=s\,,$ are non-zero.
	Substituting into \eqref{eq:ads stab eqn} , we derive the following linear system:
	\begin{itemize}
		\item $(0'\,, 1)\quad s\,B^{0'2}-m\,B^{01}=0\,;$
		\item $(0\phantom{'}\,, 1)\quad s\,B^{02}+m\,B^{0'1}=0\,;$
		\item $(0'\,, 2)\quad s\,B^{0'1}+m\,B^{02}=0\,;$
		\item $(0\phantom{'}\,, 2)\quad s\,B^{01}-m\,B^{0'2}=0\,;$
		\item $(0'\,, i)\quad B^{0i}=0\,;$
		\item $(0\phantom{'}\,, i)\quad B^{0'i}=0\,;$
		\item $(1\phantom{'}\,, i)\quad B^{2i}=0\,;$
		\item $(2\phantom{'}\,, i)\quad B^{1i}=0\,,$
	\end{itemize}
	where $i,j=3\,,\ldots\,,d-1$.
	This system of linear equations has a singularity at $m=s$,
	which necessitates a separate analysis.
	For the generic case $m\neq s$, 
	the first four equations provide the solutions
	$B^{0'1}=B^{0'2}=B^{01}=B^{02}=0$.
	Consequently, the stabiliser subalgebra is given by
	\be
		\fg^{\f}=\span{ J_{0'0}\,,\ J_{12}\,,\ J_{ij}}\,,
	\ee
	where $i,j=3\,,\ldots\,, d-1$,
	and it is isomorphic to $\fu(1)\oplus \fu(1)\oplus \fso(d-3)$.
	Since the stabiliser algebra $\fg^{\f}$ is a $\frac12(d^2-7d+16)$-dimensional,
	the corresponding coadjoint orbit $\cO^{\f}$ is a $2(2d-4)$-dimensional phase space
	which has the same dimension as that of the Poincar\'e massive spinning case.

	On the other hand, for $m=s$ case, the representative is given as
	\be
		\f'=s(\cJ^{0'0}+\cJ^{12})\,,
	\ee
	and the solutions from the first four equations
	are $B^{0'1}=-B^{02}$ and $B^{0'2}=B^{01}$.
	Therefore, the corresponding stabiliser algebra is given as
	\be
		\fg^{\f'}=\span{J_{0'1}-J_{02}\,,\ J_{0'2}+J_{01}\,,\ J_{0'0}+J_{12}\,,\ J_{0'0}-J_{12}\,,\ J_{0i}\,,\ \ J_{ij}}\,,
	\ee
	and it is isomorphic to $\fu(1,1)\oplus \fso(d-3)$.

	Note that $\dim \fg^{\f'}$ is larger than $\dim \fg^{\f}$ by two,
	so the coadjoint orbit $\cO^{\f'}$ corresponds to a smaller phase space of dimension $2(2d-5)$.
	This implies that $\f'$ is the counterpart of the Poincar\'e massless particle.
		
	In conclusion, $\f=m\,\cJ^{0'0}+s\,\cJ^{12}$ 
	can be interpreted as a massive spinning particle in AdS for $m\neq s$.
	In contrast, when $m=s$, the representative $\f'=s(\cJ^{0'0}+\cJ^{12})$
	corresponds to a massless spinning particle in AdS.
	This implies that the condition $m=s$ 
	serves as the classical analogue of the AdS massless condition. 
	\footnote{A subtlety arises concerning the relative magnitude of $m$ and $s$.
	In the context of field theory, the condition $m=s$ corresponds to
	the classical limit of the unitary bound, 
	while the region $m<s$ is typically associated with non-unitary representation.
	However, the orbit method suggests that  
	each coadjoint orbit corresponds to a unitary representation.
	This inconsistency can be resolved by
	the existence of novel unitary representations in the $m<s$ regime.
	For further discussion on this point,
	we refer the reader to \cite{part1}, and we will not elaborate on it here.}
	See \cite{Enayati:2023lld} for further discussions on the contraction of 
	AdS coadjoint orbits to their Poincar\'e counterparts.
	Accordingly, throughout the rest of the paper,
	we refer to $\f$ as the AdS massive spinning particle and 
	to $\f'$ as the AdS massless spinning particle.

\section{Worldline Actions}
\label{sec:action}

In Section \ref{sec:review}, we discussed
the formulation of manifestly covariant worldline actions
for relativistic particles in either $ISO(1,d-1)$ or $SO(2,d-1)$.
In this section, we construct the worldline action for 
Poincar\'e and AdS spinning particles
based on the representative vectors discussed in the previous section.

 	\subsection{Poincar\'e Spinning Particle}
	Let us consider the Poincar\'e massive spinning particle.
	Based on the representative given in \eqref{eq:poin spin rep}, 
	the homogeneous part corresponds to a rank-two matrix of the form,
	\be \label{eq:hom rep mat}
		\f_H=\frac12\left(	\begin{array}{cccc}
						0 & s & 0 & \cdots \\
						-s & 0 & 0 & \cdots \\
						0 & 0 & 0& \cdots \\
						{} & \vdots & {} & \ddots 
						\end{array} \right) \,,
	\ee
	where the first column corresponds to $(\f_H)^{a1}$ and 
	the second column to $(\f_H)^{a2}$.
	Note that the factor $1/2$ appearing in \eqref{eq:hom rep mat} 
	arises from the canonical pairing defined in\eqref{eq:dual space}.
	The representative in \eqref{eq:hom rep mat}, 
	can be transformed into the symplectic matrix 
	with $2\times (d+1)$ matrix $T$,
	\be\label{eq:poin t mat}\ba
		T=\frac1{\sqrt{2}}\left( 	
		\begin{array}{ccc}
			\sqrt{s} & 0 & \cdots \\
			0 & \sqrt{s} & \cdots
			\end{array} 
		\right) \,.
	\ea\ee
	From this $T$ matrix, $\vf$ are given as 
	\be
		\vf_M=m^2\,,\quad (\vf_I)_\a=0\,,\quad 
		(\vf_H)_{\a\b}=\frac12 \left( \begin{array}{cc}
					s & 0  \\
					0 & s 
				\end{array} 
		\right)\,.
	\ee
	Following the substitutions \eqref{eq:subst poin}, we set the variables for our convenience as
	\be \label{eq:poin spin var}
		p_a=m\L_{a0}\,, \ \p^a =\frac1{\sqrt s}\L^a{}_1\,,\ \c^a=\sqrt{s}\,\L^a{}_2\,,
	\ee
	and put them into \eqref{eq:inhom action} 
	then we can find the worldline action of the massive spinning particle as
	\be\ba \label{eq:massive action}
		S[x\,, p\,, \p\,, \c\,,  A_I] = \int\,\Big[\, &p_a \dd x^a + \p_a \dd \c^a 	\\
									&\qquad+A_{pp} \left( p^2+m^2 \right)+ A_{\p\p} \left(\p^2-s^2 \right) + A_{\c\c}\left( \c^2-1\right) \\
									&\qquad+A_{p\p}\left( p\cdot\p \right) +A_{p\c}\left( p\cdot\c \right) +A_{\p\c}\left( \p\cdot\c \right) 								\Big]\,.
	\ea\ee	
	Note that the constraints of action support  information of the orbit.
	The particle mass has been encoded in the mass-shell constraint $p^2+m^2\approx0$,
	and spin information is carried by two spin constraints $\p^2-s^2\approx0$ and $\c^2-1\approx1$.
	Other constraints in \eqref{eq:massive action}  give the conditions 
	between momentum and spin vectors $\p_a\,,\c^a$
	or between spin vectors $\p_a\,,\c^a$.
	Such restriction already had been discussed in 
	\cite{souriau1970structure}.
	
	Also, quantisation of \eqref{eq:massive action} will restrict 
	$s$ to integer values 
	and modify $s^2$ to $s(s+1)$. 
	We postpone further discussion to our future works.

	The six constraints have the following non-trivial Poisson brackets:
	\be
		\{\p^2\,, \p\cdot\c\}=2\p^2 \approx 2s^2\,,\ \{\c^2\,, \p\cdot\c\}=-2\c^2 \approx -2\,,\ \{p\cdot \p\,, p\cdot\c\}=p^2 \approx -m^2\,.
	\ee
	Among the six constraints, the first-class constraints are
	\be
		p^2\approx -m^2 \,, \quad \p^2+s^2\,\c^2\approx 2s^2\,,
	\ee
	satisfying Poisson bracket
	\be
		\{ p^2\,, \p^2+s^2\,\c^2 \} = 0\,.
	\ee
	This implies that the first-class constraints generate $\bbR \oplus \fu(1)$ Lie algebra.
	The embedding phase space has $4d$ degrees of freedom,
	but the constraints eliminate $8$ of them,
	resulting in a reduced phase space of dimension $2(2d-4)$-dimensional.
	Notably, this matches the dimension of the coadjoint orbit 
	associated with the representative in \eqref{eq:poin spin rep}. 

	In the massless case \eqref{eq:poin massless spin rep},
	the matrix \eqref{eq:hom rep mat} and \eqref{eq:poin t mat} remain essentially the same as in the massive case.
	As a result, the components of $\vf$ are expressed in nearly identical manner,
	except for $\vf_M$, which is modified as follows:
	\be
		\vf_M=0\,,\quad (\vf_I)_\a=0\,,\quad 
		(\vf_H)_{\a\b}=\frac12 \left( \begin{array}{cc}
					s & 0  \\
					0 & s 
				\end{array} 
		\right)\,.
	\ee
	By adopting the same variables in \eqref{eq:poin spin var},
	we obtain the worldline action for the massless spinning particle as follows:
	\be\ba \label{eq:massless action}
		S[x\,, p\,, \p\,, \c\,,  A_I] = \int\,\Big[\, &p_a \dd x^a + \p_a \dd \c^a 	\\
							&\qquad+A_{pp} \left( p^2 \right)+ A_{\p\p} \left(\p^2-s^2 \right) + A_{\c\c}\left( \c^2-1\right) \\
							&\qquad+A_{p\p}\left( p\cdot\p \right) +A_{p\c}\left( p\cdot\c \right)	 +A_{\p\c}\left( \p\cdot\c \right) 	\Big]\,.
	\ea\ee	
	The difference between \eqref{eq:massive action} and \eqref{eq:massless action}
	lies in the vanishing of $m^2$ in the latter.
	Furthermore, the parameter $E$ from the representative vector \eqref{eq:poin massless spin rep}
	does not appear in the action \eqref{eq:massless action},
	whereas the mass parameter $m$ of the massive particle \eqref{eq:poin spin rep}
	explicitly enters the action.
	This implies that $E$ is a nilpotent parameter that can be freely rescaled
	via the coadjoint action of $J^{+-}$,
	which acts as a scaling transformation on $E$.
	
	Among the six constraints in the massless particle action \eqref{eq:massless action},
	the non-trivial Poisson brackets are given by
	\be
		\{\p^2\,, \p\cdot\c\}=2\p^2 \approx 2s^2\,,\ \{\c^2\,, \p\cdot\c\}=-2\c^2 \approx -2\,,\ \{p\cdot \p\,, p\cdot\c\}=p^2 \approx 0\,,
	\ee
	and the first-class constraints are given by
	\be
		p^2\approx 0 \,, \quad \p^2+s^2\,\c^2\approx 2s^2\,, \quad p\cdot\p\approx0\,,\quad p\cdot\c\approx0\,,
	\ee
	which they satisfy the Poisson brackets relation
	\be\bg
		\{p\cdot\p\,,p\cdot\c\}=p^2\,,\quad \{p\cdot\p\,,\p^2+s^2\,\c^2\}=2s^2\,p\cdot\c\,,\quad \{p\cdot\c\,,\p^2+s^2\,\c^2\}=2s^2\,p\cdot\p\,, \\
		\{p^2\,,\p^2+s^2\,\c^2\}=\{p^2\,,p\cdot\p\}=\{p^2\,,p\cdot\c\}=0\,.
	\eg\ee
	The first-class constraints form a $\fu(1)\niplus\fheis_2$ algebra,
	eliminating 10 degrees of freedom from the embedding phase space.
	Consequently, the physical degrees of freedom have dimension $2(2d-5)$.
	
	The construction of the Poincar\'e particle action
	from the coadjoint orbit has also been discussed in various literature, e.g.,
	\cite{Lyakhovich:1998ij,Lyakhovich:2000zy,Andrzejewski:2020qxt,part1}.
	
	Remark that, in both massive and massless cases, 
	the Lie algebra associated with the first-class constraints
	coincides with the stabiliser algebra, modulo the $d$-dependent parts.
	This observation implies a deep correspondence between the coadjoint orbit and the constraint structure,
	a relation known as the dual pair correspondence \cite{MR0986027,MR0985172}.
	See also \cite{Basile:2020gqi} for the review of this.
	
	\subsection{AdS Spinning Particle}
	
	The AdS spinning particle corresponds to the coadjoint orbit
	associated with the representative vector in \eqref{eq:adsparticle}.
	This representative vector \eqref{eq:adsparticle} can be expressed as a $(d+1)\times (d+1)$ matrix
	\be
		\f=\frac12\left(\begin{array}{cccccc}
					0 & m & 0 & 0 & 0 & \cdots \\
					-m & 0 & 0 & 0 & 0 & \cdots \\
					0 & 0 & 0 & s & 0 & \cdots \\
					0 & 0 & -s & 0 & 0 & \cdots \\
					{} & {} & \vdots & {} & {} & \ddots
				     	\end{array}
				\right)\,.
	\ee
	Since the rank of $\f$ is 4, 
	$\f$ can be transformed from the $4\times4$ symplectic matrix $\O$ 
	with a $4\times (d+1)$ transformation matrix,
	\be
		T=\frac1{\sqrt2}\left(\begin{array}{ccccc}
						\sqrt{m} & 0 & 0 & 0 & \cdots \\
						0 & \sqrt{m} & 0 & 0 & \cdots \\
						0 & 0 & \sqrt{s} & 0 & \cdots \\ 
						0 & 0 & 0 & \sqrt{s} & \cdots 
						\end{array}
					\right)\,.
	\ee
	From the above, we find $\vf$ 
	\be \label{eq:ads particle const}
		\vf=\frac12\left(\begin{array}{cccc}
						-m & 0 & 0 & 0 \\
						0 & -m & 0 & 0 \\
						0 & 0 & s & 0   \\ 
						0 & 0 & 0 & s  
						\end{array}
				\right)\,,
	\ee
	which appears as a constant piece of the constraints.
	From \eqref{eq:redef var},
	we substitute $X\in SO(2,d-1)$, as follows
	\be\label{eq:AdS redef}
		X^A=\sqrt{m}\, X^A{}_{0'}\,,\quad P^A=\frac1{\sqrt m}X^A{}_{0}\,,\quad \c^A={\sqrt s}\,X^A{}_{1}\,,\quad \p^A=\frac1{\sqrt s}X^A{}_{2}\,,
	\ee
	where $A,B=0'\,,0\,,1\,,\ldots\,,d-1$ are indices for ambient space.
	Implementing \eqref{eq:ads particle const} and \eqref{eq:AdS redef} into \eqref{eq:hom action},
	the manifestly covariant worldline action for the AdS spinning particle is given by
	\be\ba \label{ads mass action}
		S[X\,, P\,, \p\,, \c\,,  A_I] = \int\,\Big[ &P_A \dd X^A + \p_A\dd\c^A +A_{PP} \left(P^2+m^2 \right)+A_{XX}\left(X^2+1 \right) \\
										&\qquad+ A_{\p\p} \left(\p^2-s^2 \right) + A_{\c\c}\left( \c^2-1\right) \\
										&\qquad+A_{PX}\left(P\cdot X \right)+A_{P\p}\left( P\cdot\p \right) +A_{p\c}\left( P\cdot\c \right) \\
										&\qquad+A_{X\p}\left(X\cdot \p \right)+A_{X\c}\left(X\cdot\c\right) +A_{\p\c}\left( \p\cdot\c \right) \Big]\,.
	\ea\ee
	Note that the constraint $P^2\approx m^2$ can be interpreted as the mass-shell constraint,
	while $P\cdot X\approx 0$ serves as a homogeneity condition.
	The $X^2\approx-1$ indicates that $X$ lies on the ambient space of AdS spacetime.
	The AdS particle actions have also been discussed in \cite{Kuzenko:1994vh,Kuzenko:1995aq,Ghosh:2005se}.
	Similarly to the Poincar\'e case in  \eqref{eq:massive action},
	these constraints carry the information of the particle.
	As discussed in Sec. \ref{subsec:ads orbit}, 
	only integer $m$ and $s$ values in \eqref{ads mass action} 
	will be allowed after quantisation.

	There are eight non-trivial Poisson brackets between the constraints:
	\be \bg
		\{P^2\,, P\cdot X\}=2P^2 \approx -2m^2 \,, \ \{X^2\,, P\cdot X\}= -2X^2 \approx 2 \,,\\
		\{\p^2\,, \p\cdot\c\} = 2\p^2 \approx 2s^2 \,, \ \{\c^2\,, \p\cdot\c\}=-2\c^2 \approx -2\,, \\
		\{P\cdot\p\,, P\cdot\c \} = P^2 \approx -m^2 \,,\ \{P\cdot\p\,, X\cdot\p\}= \p^2 \approx s^2 \,,\\
		\{X\cdot\p\,, X\cdot\c \} = X^2 \approx -1 \,, \ \{P\cdot\c\,, X\cdot\c\} = \c^2 \approx 1\,.
	\eg \ee
	Among the constraints, there are two first-class constraints given as 	
	\be\label{eq:ads first class const}
		P^2+X^2\,, \quad \p^2+\c^2\,,
	\ee
	for any $m$ and $s$, and two additional first-class constraints appear
	\be\label{eq:add first class const}
		P\cdot\c+X\cdot\p\,,\quad X\cdot\c-P\cdot\p\,,
	\ee
	only if $s=m$. 
	As discussed in Section \ref{subsec:ads orbit}, 
	the condition $m=s$ corresponds to the massless condition of the AdS particle. 
	The embedding phase space possesses $4(d+1)$ degrees of freedom.
	For the massive case, the imposition of constraints reduces this to $2(2d-4)$ physical degrees of freedom.
	On the other hand, the massless case involves 
	two additional  first-class constraints, given in \eqref{eq:add first class const},
	which further reduce the physical degrees of freedom to $2(2d-5)$.

	For the massive case, the first-class constraints 
	\eqref{eq:ads first class const} satisfy the Poisson bracket,
	\be
		\{P^2+X^2\,,\p^2+\c^2\}=0\,,
	\ee
	so they generate $\fu(1)\oplus\fu(1)$ algebra.
	In the case of the massless particle, 
	\eqref{eq:ads first class const} and \eqref{eq:add first class const} satisfy
	\be\bg
		\{P^2+X^2\,,P\cdot\c+X\cdot\p\}=2(P\cdot\p-X\cdot\c)\,,\ 
		\{P^2+X^2\,,X\cdot\c-P\cdot\p\}=2(P\cdot\c+X\cdot\p)\,, \\ 
		\{\p^2+\c^2\,,P\cdot\c+X\cdot\p\}=2(P\cdot\p-X\cdot\c)\,,\  
		\{\p^2+\c^2\,,X\cdot\c-P\cdot\p\}=2(X\cdot\p+P\cdot\c)\,,\\ 
		\{P^2+X^2\,,\p^2+\c^2\}=0\,,\ \{P\cdot\c+X\cdot\p\,,X\cdot\c-P\cdot\p\}=P^2+X^2+\c^2+\p^2\,,
	\eg\ee
	and  these constraints form a $\fu(1,1)$ algebra. 
	
	Analogously to the Poincaré case,
	the Lie algebras formed from the first-class constraints 
	are identical to the stabiliser algebras
	without $d$-dependent components. 
	These facts also imply the dual pair correspondence\cite{part1, Basile:2020gqi}.

\section{Conclusion}
\label{sec:conclusion}

We have discussed the construction of manifestly covariant worldline actions
using the orbit method by introducing Hamiltonian constraints,
enabling  the formulation of \textit{ab initio} worldline actions.
For Poincar\'e particles, we chose inhomogeneous representative vectors
corresponding to the mass type of the particle,
and selected the homogeneous representative vectors from the scalar particle stabiliser subalgebra.
For AdS particles, we chose the representatives that can be viewed as analogues
of the Poincar\'e massive particles. 
Remarkably, when the mass parameter equals the spin,
the coadjoint orbit becomes smaller,
which can be interpreted as the phase space of an AdS massless particle.
From the representative vector, one can readily determine the stabiliser algebra 
and thereby identify the geometry of the associated coadjoint orbit. 

Starting from these representatives,
worldline actions can be systematically constructed.
A particularly efficient and geometrically motivated approach is
to impose the defining condition of the isometry as a Hamiltonian constraint,
which leads to a manifestly covariant formulation of the worldline action.
By comparing the stabiliser subalgebra with the  first-class constraint, 
we observe not only a matching in the number of degrees of freedom
between the coadjoint orbit and the physical phase space,
but also a structural equivalence between the stabiliser algebra
and the set of first-class constraints dependent up to dimension-dependent parts
such as $\fso(d-n)$ or $\fiso(d-n)$.
This correspondence offers a hint of the dual pair correspondence,
implying a one-to-one correspondence 
between coadjoint orbits of the isometry group and Hamiltonian constraints.

The Hamiltonian constraints in the manifestly covariant actions
possess the information of the particles
which was transferred from the representatives of the orbits,
and some of orbit components should be taken integer values after the quantisation.
The quantization of the actions will not only permit the integer values of certain particle information,
but also provide unitary irreducible representations of the isometries.
For quantizing the action, well-known machineries, 
such as path-integral formulation or BRST quantization,
can be utilised.

This framework is flexible and can be extended to various particle species,
including those in flat and AdS spacetimes, as well as
to models with different symmetry groups
such as dS, twistor groups, Galilean symmetry, and supergroups.
Furthermore, we hope that this work sheds new light on the various physics topics
making use of worldline action.

\acknowledgments
I am grateful for helpful discussions with Euihun Joung
and also thanks to Sejin Kim and Minkyeu Cho for discussions about the draft. 
The authors declare no competing financial interest.
The author was supported by the National Research Foundation of Korea (NRF) grant funded by the Korea government (MSIT) (No. 2022R1F1A1074977).
Also, beneficiary of an individual grant from QUC supported by KIAS Individual Grant (QP099501)
via the Quantum Universe Center at Korea Institute for Advanced Study.

\appendix

\section{Transformation of an Anti-symmetric Matrix into a Symplectic Matrix}
\label{sec:proof}

Suppose $\f$ is an $N\times N$ anti-symmetric matrix with real entries. 
Then $\f^T\,\f$ is a symmetric matrix, so it must have positive or null eigenvalues:
\be\ba
	\f^T\f\, v_i = \l_i^2 v_i \,, \\
	\f^T\f\, \k_a = 0\,,
\ea\ee
where $i=1\,,\cdots\,,M=\rank \f$, 
We assume that $\l_i$ are positive, without loss of generality.
Let us define new $M$ vectors $w_i$ as
\be
	w_i=\frac1{\l_i}\f\, v_i\,.
\ee
Since $\f$ is an anti-symmetric matrix,
$w_i$ is also an eigenvector of $\f^T\f$ 
with the eigenvalue $\l_i^2$,
as follows
\be
	\f^T\f\, w_i = \frac 1{\l_i} \f^T \f \f\,v_i = -\l_i \f^T \, v_i = \l_i^2 w_i\,.
\ee
But one can check that $v_i$ and $w_i$ are degenerate eigenvectors
\be \ba
	v_i\cdot w_i 	&= \left(-\frac 1{\l_i} \f\, w_i \right) \cdot \left ( \frac1{\l_i}\f\, v_i \right) \\
				&= -\frac 1{\l_i^2}\, w_i^T\, \f^T\f\, v_i \\
				&= -w_i \cdot v_i \,.
\ea \ee
Thus, if we choose the matrix $Q$ as
\be
	Q = \left( \frac 1{\sqrt{\l_1}}\vec v_1\,,\ \frac 1{\sqrt{\l_1}}\vec w_1\,,\ \cdots
	\frac 1{\sqrt{\l_M}}\vec v_M\,,\ \frac 1{\sqrt{\l_M}}\vec w_M\,,\  \vec{\k}_1 \,, \cdots \vec{\k}_{N-2M} \right)
\ee
then it satisfies
\be
	Q^T\f\, Q = \O\,.
\ee
By construction, $Q$ is a non-degenerate matrix, 
so it can be inverted easily.
Then, we can find $T$ matrix which satisfies
\be
	T^T\, \O\, T = \f \,.
\ee

\bibliographystyle{JHEP}
\bibliography{biblio}

\providecommand{\href}[2]{#2}\begingroup\raggedright\begin{thebibliography}{10}

\bibitem{Frenkel:1926zz}
J.~Frenkel, \emph{{Die Elektrodynamik des rotierenden Elektrons}},
  \href{http://dx.doi.org/10.1007/BF01397099}{\emph{Z. Phys.} {\bf 37} (1926)
  243--262}.

\bibitem{Gershun:1979fb}
V.~D. Gershun and V.~I. Tkach, \emph{{CLASSICAL AND QUANTUM DYNAMICS OF
  PARTICLES WITH ARBITRARY SPIN}}, {\emph{JETP Lett.} {\bf 29} (1979)
  288--291}.

\bibitem{Casalbuoni:1976tz}
R.~Casalbuoni, \emph{{The Classical Mechanics for Bose-Fermi Systems}},
  \href{http://dx.doi.org/10.1007/BF02729860}{\emph{Nuovo Cim. A} {\bf 33}
  (1976) 389}.

\bibitem{Berezin:1976eg}
F.~A. Berezin and M.~S. Marinov, \emph{{Particle Spin Dynamics as the Grassmann
  Variant of Classical Mechanics}},
  \href{http://dx.doi.org/10.1016/0003-4916(77)90335-9}{\emph{Annals Phys.}
  {\bf 104} (1977) 336}.

\bibitem{Brink:1976sz}
L.~Brink, S.~Deser, B.~Zumino, P.~Di~Vecchia and P.~S. Howe, \emph{{Local
  Supersymmetry for Spinning Particles}},
  \href{http://dx.doi.org/10.1016/0370-2693(76)90115-5}{\emph{Phys. Lett. B}
  {\bf 64} (1976) 435}.

\bibitem{Barducci:1976qu}
A.~Barducci, R.~Casalbuoni and L.~Lusanna, \emph{{Supersymmetries and the
  Pseudoclassical Relativistic electron}},
  \href{http://dx.doi.org/10.1007/BF02730291}{\emph{Nuovo Cim. A} {\bf 35}
  (1976) 377}.

\bibitem{part1}
T.~Basile, E.~Joung and T.~Oh, \emph{{Manifestly covariant worldline actions
  from coadjoint orbits. Part I. Generalities and vectorial descriptions}},
  \href{http://dx.doi.org/10.1007/JHEP01(2024)018}{\emph{JHEP} {\bf 01} (2024)
  018}, [\href{http://arxiv.org/abs/2307.13644}{{\tt 2307.13644}}].

\bibitem{Bastianelli:2008vh}
F.~Bastianelli, O.~Corradini, P.~A.~G. Pisani and C.~Schubert, \emph{{Scalar
  heat kernel with boundary in the worldline formalism}},
  \href{http://dx.doi.org/10.1088/1126-6708/2008/10/095}{\emph{JHEP} {\bf 10}
  (2008) 095}, [\href{http://arxiv.org/abs/0809.0652}{{\tt 0809.0652}}].

\bibitem{Bastianelli:2023oyz}
F.~Bastianelli and M.~D. Paciarini, \emph{{Worldline path integrals for the
  graviton}}, \href{http://dx.doi.org/10.1088/1361-6382/ad3f69}{\emph{Class.
  Quant. Grav.} {\bf 41} (2024) 115002},
  [\href{http://arxiv.org/abs/2305.06650}{{\tt 2305.06650}}].

\bibitem{Comberiati:2022cpm}
F.~Comberiati and C.~Shi, \emph{{Classical Double Copy of Spinning Worldline
  Quantum Field Theory}},
  \href{http://dx.doi.org/10.1007/JHEP04(2023)008}{\emph{JHEP} {\bf 04} (2023)
  008}, [\href{http://arxiv.org/abs/2212.13855}{{\tt 2212.13855}}].

\bibitem{Rempel:2015foa}
T.~Rempel and L.~Freidel, \emph{{Interaction Vertex for Classical Spinning
  Particles}}, \href{http://dx.doi.org/10.1103/PhysRevD.94.044011}{\emph{Phys.
  Rev. D} {\bf 94} (2016) 044011}, [\href{http://arxiv.org/abs/1507.05826}{{\tt
  1507.05826}}].

\bibitem{Schubert:2023bed}
C.~Schubert, \emph{{The worldline formalism in strong-field QED}},
  \href{http://dx.doi.org/10.1088/1742-6596/2494/1/012020}{\emph{J. Phys. Conf.
  Ser.} {\bf 2494} (2023) 012020}, [\href{http://arxiv.org/abs/2304.07404}{{\tt
  2304.07404}}].

\bibitem{kirillov1962unitary}
A.~A. Kirillov, \emph{Unitary representations of nilpotent lie groups},
  {\emph{Russian mathematical surveys} {\bf 17} (1962) 53}.

\bibitem{kirillov2004lectures}
A.~A. Kirillov, \emph{Lectures on the orbit method}, vol.~64.
\newblock American Mathematical Soc., 2004.

\bibitem{auslander1967quantization}
L.~Auslander and B.~Kostant, \emph{Quantization and representations of solvable
  lie groups}, .

\bibitem{Kostant:1969zz}
B.~Kostant, \emph{{On certain unitary representations which arise from a
  quantization theory}},
  \href{http://dx.doi.org/10.1007/3-540-05310-7_28}{\emph{Conf. Proc. C} {\bf
  690722} (1969) 237--253}.

\bibitem{auslander1971polarization}
L.~Auslander and B.~Kostant, \emph{Polarization and unitary representations of
  solvable lie groups}, {\emph{Inventiones mathematicae} {\bf 14} (1971)
  255--354}.

\bibitem{souriau1970structure}
J.-M. Souriau, \emph{Structure des syst\`emes dynamiques: ma\^{i}trises de
  math\'ematiques}.
\newblock 1970.

\bibitem{Duval:2014ppa}
C.~Duval and P.~A. Horvathy, \emph{{Chiral fermions as classical massless
  spinning particles}},
  \href{http://dx.doi.org/10.1103/PhysRevD.91.045013}{\emph{Phys. Rev. D} {\bf
  91} (2015) 045013}, [\href{http://arxiv.org/abs/1406.0718}{{\tt 1406.0718}}].

\bibitem{Andrzejewski:2020qxt}
K.~Andrzejewski, C.~Gonera, J.~Gonera, P.~Kosinski and P.~Maslanka,
  \emph{{Spinning particles, coadjoint orbits and Hamiltonian formalism}},
  \href{http://dx.doi.org/10.1016/j.nuclphysb.2022.115664}{\emph{Nucl. Phys. B}
  {\bf 975} (2022) 115664}, [\href{http://arxiv.org/abs/2008.09478}{{\tt
  2008.09478}}].

\bibitem{kirillov2012elements}
A.~A. Kirillov, \emph{Elements of the Theory of Representations}, vol.~220.
\newblock Springer Science \& Business Media, 2012.

\bibitem{vogan1998method}
D.~A. Vogan~Jr, \emph{The method of coadjoint orbits for real reductive
  groups}, {\emph{Representation theory of Lie groups (Park City, UT, 1998)}
  {\bf 8} (1998) 179--238}.

\bibitem{Oblak:2016eij}
B.~Oblak, \emph{{BMS Particles in Three Dimensions}}.
\newblock PhD thesis, U. Brussels, Brussels U., 2016.
\newblock \href{http://arxiv.org/abs/1610.08526}{{\tt 1610.08526}}.
\newblock 10.1007/978-3-319-61878-4.

\bibitem{Wigner:1939cj}
E.~P. Wigner, \emph{{On Unitary Representations of the Inhomogeneous Lorentz
  Group}}, \href{http://dx.doi.org/10.2307/1968551}{\emph{Annals Math.} {\bf
  40} (1939) 149--204}.

\bibitem{Bargmann:1948ck}
V.~Bargmann and E.~P. Wigner, \emph{{Group Theoretical Discussion of
  Relativistic Wave Equations}},
  \href{http://dx.doi.org/10.1073/pnas.34.5.211}{\emph{Proc. Nat. Acad. Sci.}
  {\bf 34} (1948) 211}.

\bibitem{Lyakhovich:1996we}
S.~L. Lyakhovich, A.~Y. Segal and A.~A. Sharapov, \emph{{A Universal model of D
  = 4 spinning particle}},
  \href{http://dx.doi.org/10.1103/PhysRevD.54.5223}{\emph{Phys. Rev. D} {\bf
  54} (1996) 5223--5238}, [\href{http://arxiv.org/abs/hep-th/9603174}{{\tt
  hep-th/9603174}}].

\bibitem{Kuzenko:1994ju}
S.~M. Kuzenko, S.~L. Lyakhovich and A.~Y. Segal, \emph{{A Geometric model of
  arbitrary spin massive particle}},
  \href{http://dx.doi.org/10.1142/S0217751X95000735}{\emph{Int. J. Mod. Phys.
  A} {\bf 10} (1995) 1529--1552},
  [\href{http://arxiv.org/abs/hep-th/9403196}{{\tt hep-th/9403196}}].

\bibitem{Rawnsley1975}
J.~H. Rawnsley, \emph{{Representations of a semi-direct product by
  quantization}},
  \href{http://dx.doi.org/10.1017/S0305004100051793}{\emph{Mathematical
  Proceedings of the Cambridge Philosophical Society} {\bf 78} (1975)
  345--350}.

\bibitem{Carinena:1989uw}
J.~F. Carinena, J.~M. Gracia-Bondia and J.~C. Varilly, \emph{{Relativistic
  Quantum Kinematics in the Moyal Representation}},
  \href{http://dx.doi.org/10.1088/0305-4470/23/6/015}{\emph{J. Phys. A} {\bf
  23} (1990) 901}.

\bibitem{Baguis1998}
P.~Baguis, \emph{{Semidirect products and the Pukanszky condition}},
  \href{http://dx.doi.org/10.1016/S0393-0440(97)00028-4}{\emph{Journal of
  Geometry and physics} {\bf 25} (1998) 245--270},
  [\href{http://arxiv.org/abs/dg-ga/9705005}{{\tt dg-ga/9705005}}].

\bibitem{Cushman_2006}
R.~Cushman and W.~van~der Kallen, \emph{Adjoint and coadjoint orbits of the
  poincar{\'{e}} group},
  \href{http://dx.doi.org/10.1007/s10440-006-9031-8}{\emph{Acta Applicandae
  Mathematicae} {\bf 90} (may, 2006) 65--89}.

\bibitem{hudon2010coadjoint}
V.~Hudon and S.~T. Ali, \emph{Coadjoint orbits of the poincar\'e group in 2+1
  dimensions and their coherent states},  2010.

\bibitem{Gracia-Bondia:2017fai}
J.~M. Gracia-Bondia, F.~Lizzi, J.~C. Varilly and P.~Vitale, \emph{{The Kirillov
  picture for the Wigner particle}},
  \href{http://dx.doi.org/10.1088/1751-8121/aac3b3}{\emph{J. Phys. A} {\bf 51}
  (2018) 255203}, [\href{http://arxiv.org/abs/1711.09608}{{\tt 1711.09608}}].

\bibitem{Havlicek:2018tfp}
M.~Havl\'\i{}\v{c}ek, J.~Kotrbat\'y, P.~Moylan and S.~Po\v{s}ta,
  \emph{{Construction of representations of Poincar\'e group using Lie
  fields}}, \href{http://dx.doi.org/10.1063/1.4993153}{\emph{J. Math. Phys.}
  {\bf 59} (2018) 021702}.

\bibitem{Lahlali:2021nrf}
I.~A. Lahlali, N.~Boulanger and A.~Campoleoni, \emph{{Coadjoint Orbits of the
  Poincar\'e Group for Discrete-Spin Particles in Any Dimension}},
  \href{http://dx.doi.org/10.3390/sym13091749}{\emph{Symmetry} {\bf 13} (2021)
  1749}.

\bibitem{Enayati:2023lld}
M.~Enayati, J.-P. Gazeau, M.~A. del Olmo and H.~Pejhan, \emph{{Anti-de
  Sitterian {\textquotedblleft}massive{\textquotedblright} elementary systems
  and their Minkowskian and Newton-Hooke contraction limits}},
  \href{http://dx.doi.org/10.1063/5.0168115}{\emph{J. Math. Phys.} {\bf 66}
  (2025) 053501}, [\href{http://arxiv.org/abs/2307.06690}{{\tt 2307.06690}}].

\bibitem{Lyakhovich:1998ij}
S.~L. Lyakhovich, A.~A. Sharapov and K.~M. Shekhter, \emph{{Massive spinning
  particle in any dimension. 1. Integer spins}},
  \href{http://dx.doi.org/10.1016/S0550-3213(98)00617-8}{\emph{Nucl. Phys. B}
  {\bf 537} (1999) 640--652}, [\href{http://arxiv.org/abs/hep-th/9805020}{{\tt
  hep-th/9805020}}].

\bibitem{Lyakhovich:2000zy}
S.~L. Lyakhovich, A.~A. Sharapov and K.~M. Shekhter, \emph{{A Uniform model of
  the massive spinning particle in any dimension}},
  \href{http://dx.doi.org/10.1016/S0217-751X(00)00211-X}{\emph{Int. J. Mod.
  Phys. A} {\bf 15} (2000) 4287--4300},
  [\href{http://arxiv.org/abs/hep-th/0002247}{{\tt hep-th/0002247}}].

\bibitem{MR0986027}
R.~Howe, \emph{Remarks on classical invariant theory},
  \href{http://dx.doi.org/10.2307/2001418}{\emph{Trans. Amer. Math. Soc.} {\bf
  313} (1989) 539--570}.

\bibitem{MR0985172}
R.~Howe, \emph{Transcending classical invariant theory},
  \href{http://dx.doi.org/10.2307/1990942}{\emph{J. Amer. Math. Soc.} {\bf 2}
  (1989) 535--552}.

\bibitem{Basile:2020gqi}
T.~Basile, E.~Joung, K.~Mkrtchyan and M.~Mojaza, \emph{{Dual Pair
  Correspondence in Physics: Oscillator Realizations and Representations}},
  \href{http://dx.doi.org/10.1007/JHEP09(2020)020}{\emph{JHEP} {\bf 09} (2020)
  020}, [\href{http://arxiv.org/abs/2006.07102}{{\tt 2006.07102}}].

\bibitem{Kuzenko:1994vh}
S.~M. Kuzenko, S.~L. Lyakhovich, A.~Y. Segal and A.~A. Sharapov, \emph{{Anti-de
  Sitter spinning particle and two sphere}},
  \href{http://arxiv.org/abs/hep-th/9411162}{{\tt hep-th/9411162}}.

\bibitem{Kuzenko:1995aq}
S.~M. Kuzenko, S.~L. Lyakhovich, A.~Y. Segal and A.~A. Sharapov, \emph{{Massive
  spinning particle on anti-de Sitter space}},
  \href{http://dx.doi.org/10.1142/S0217751X96001589}{\emph{Int. J. Mod. Phys.
  A} {\bf 11} (1996) 3307--3330},
  [\href{http://arxiv.org/abs/hep-th/9509062}{{\tt hep-th/9509062}}].

\bibitem{Ghosh:2005se}
S.~Ghosh, \emph{{The AdS particle}},
  \href{http://dx.doi.org/10.1016/j.physletb.2005.07.055}{\emph{Phys. Lett. B}
  {\bf 623} (2005) 251--257}, [\href{http://arxiv.org/abs/hep-th/0506084}{{\tt
  hep-th/0506084}}].

\end{thebibliography}\endgroup

\end{document}